# Uniaxial Compression of Suspended Single and Multilayer graphenes

A. P. Sgouros[1, 2, 3], G. Kalosakas[1, 3, 4], C. Galiotis[1, 5] and K. Papagelis[1, 3]

[1]*Institute of Chemical Engineering Sciences - Foundation of Research and Technology Hellas (FORTH / ICE-HT), GR-26504 Patras, Greece*

[2]*School of Chemical Engineering, National Technical University of Athens (NTUA), GR-15780 Athens, Greece*

[3]*Department of Materials Science, University of Patras, GR-26504 Patras, Greece*

[4]*Crete Center for Quantum Complexity and Nanotechnology (CCQCN), Physics Department, University Of Crete GR-71003 Heraklion, Greece*

[5]*School of Chemical Engineering, University of Patras, GR-26504 Patras, Greece*

**Abstract**

The mechanical response of single and multiple graphene sheets under uniaxial compressive loads was studied with molecular dynamics simulations, using different semi-empirical force fields at different boundary conditions or constrains. Compressive stress-strain curves were obtained and the critical stress/strain values were derived. For single layer graphenes, the critical stress/strain for buckling was found to scale to the inverse length square. For multilayer graphenes the critical buckling stress also decreased with increasing length, though at a slower rate than expected from elastic buckling analysis. The molecular dynamics results are compared to the linear elasticity continuum theory for loaded slabs. Qualitatively similar behavior is observed between the theory and numerical simulations for single layer graphene, while discrepancies were noted for multilayers may be due to their discrete nature and stacking interactions.

# I. Introduction

Graphene is an ultrathin membrane of atomic thickness that possesses exceptional electronic [1,2], thermal [3–5] and mechanical properties [6] which far exceed those of conventional 3D materials. Graphene being the first two-dimensional material discovered [7], due to its unique properties could form a basis for numerous applications in several branches of technology, such as in polymer matrix composite materials [8–11], optoelectronics [12,13], drug delivery [14–16], super capacitors [17] and many more. By applying stress or strain on graphene (the so called "strain engineering") one can modify its electronic properties [18], and this can make an impact to numerous applications, ranging from strain sensors [19,20] to electric circuits made from graphene [21]. Graphene ribbons under compressive loads fail by the formation of wrinkles, the wavelength, amplitude and direction of which can be controlled by modifying the boundary conditions along their edges [22]. Those wrinkles can affect drastically the local fields and the electronic properties by inducing effective magnetic fields [23,24].

The experimental Young's modulus and tensional stress of graphene have been estimated to 1 TPa and 130 GPa, respectively, by nanoindenting suspended graphene sheets with an AFM tip [6]. Another method for inducing tensile (compressive) strain to graphene is by placing graphene flakes on plastic beams [11,25–27], and then, by bending the beams [28], strains up to $\varepsilon_t$~1.5% ($\varepsilon_c$=0.7%) have been achieved [29]. Freely graphene of monoatomic thickness suspended in air is expected to have extremely low resistance to compressive loading. However, the usage of substrate can drastically change the mechanical properties of graphene, since the critical buckling strain of supported graphenes can be orders of magnitude higher than suspended graphenes [11]. This behavior is indeed very important since it confirms the role of graphene as a reinforcing agent in composite applications.

The mechanical behavior under uniaxial compression of single layer graphene (SLG) has been investigated in a few theoretical works, using molecular mechanics [30], molecular dynamics (MD) simulations [31–33] and continuum mechanics [34]. The critical buckling stress/strain of SLG has been found to be inversely proportional to the square of its length [30,32,33], in agreement with the linear elasticity theory of loaded slabs [35]. Additionally it has been found [30,33] that the critical buckling strain is the same for uniaxial compression along either the zigzag or the armchair directions. Similar theoretical studies of uniaxial compression on multilayer graphenes (MLG) are limited; a finite element approach has been used for the buckling instability of bilayer graphene [36].

In this work we performed MD simulations in order to study the mechanical response of suspended SLG and MLG under constant compressive loads (stress controlled simulations). Different semi-empirical force fields are used in order to verify the robustness of the observed behavior. The main results obtained from our simulations were the compressive stress-strain curves. The critical buckling stress ($\sigma_{crit}$) and strain ($\varepsilon_{crit}$), the Young's modulus ($E$) and the Poisson's ratio ($v$) were calculated. The critical buckling values presented here concern intrinsic properties of suspended graphenes. As mentioned above, the corresponding values for graphenes embedded in matrices, or supported on beams are expected to be significantly larger. The results of the simulations were compared with the linear elasticity theory for wide plates with unsupported sides under uniaxial compressive forces. We find that the continuum elasticity theory fails to describe the behavior observed in MLGs.

**II.  Methods**

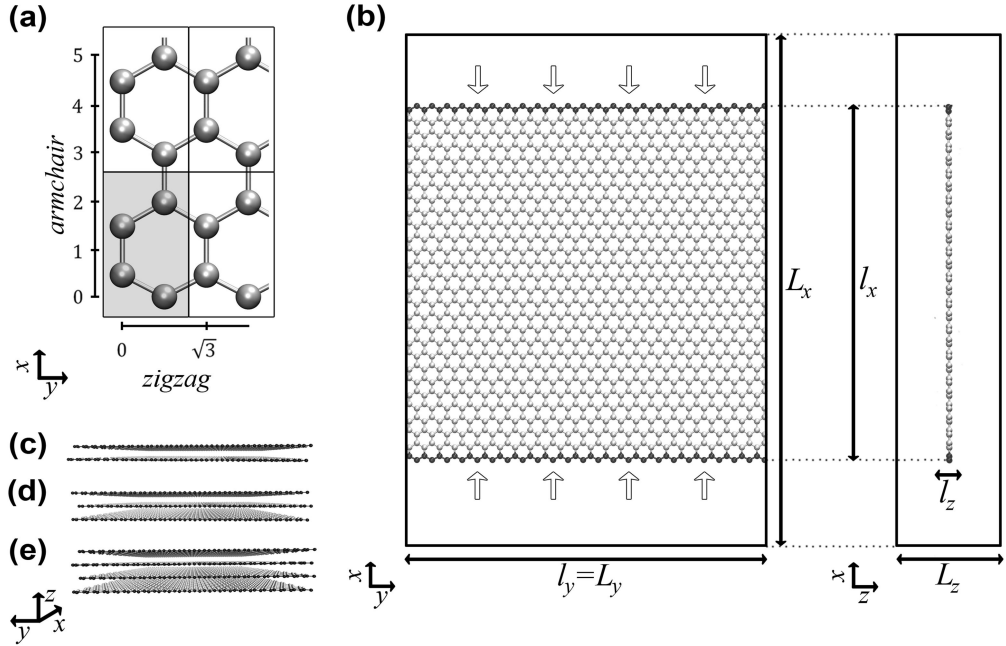

**Figure 1. a)** A 4-atom orthogonal cell replicated 2 times along the *zigzag* and *armchair* directions. The displayed lengths are in units of C-C bond lengths. **b)** A single graphene layer that is periodic only along the *y-axis* with constant compressive loads applied to the edges of the graphene at the *x* direction (black atoms). The graphene layer from (b) replicated multiple times along the *z-axis*, resulting in **(c)** bilayer, **(d)** trilayer and **(e)** quad-layer graphene.

The molecular dynamics simulations were performed with the LAMMPS package [37,38] in the isobaric-isothermal ensemble using periodic boundary conditions (PBC) and a time step equal to 1 fs. In this study we present extensive results for SLGs and MLGs (up to six layers) for various lengths at low temperatures. The temperature (pressure) of the system was maintained constant to 1K (0 bar) using the Nosé-Hoover thermostat (barostat) [39,40] with an effective relaxation time equal to 0.1 ps (1 ps). Simulations have also been performed at room temperature and similar results were found; however, the data showed a considerable scatter since the thermal fluctuations are comparable to the applied mechanical elastic energy and longer relaxation times are normally needed. In order to obtain force field independent results, calculations were performed for SLGs with three semi-empirical potentials, the Tersoff [41,42], the REBO [43] and the LCBOP I [44]. Since we found qualitatively similar behavior for these potentials concerning the length dependence of the critical buckling values, for the computationally more intensive case of MLGs we used the LCBOP potential, though some example cases were checked with the AIREBO [45] potential as well. It has been shown that the LCBOP potential provides a relatively accurate overall description of the phonon dispersion in graphene [46]. The atomistic representations and the data visualization were aided with relevant software [47,48].

In order to design the graphene sheets, the 4-atom orthogonal cell in figure **1a** was duplicated several times along the armchair and zigzag directions. Figure **1b** displays a SLG with dimensions $l_x$ (length), $l_y$ (width) and $l_z$ (thickness) inside a simulation box with dimensions $L_x$, $L_y$ and $L_z$, where $l_x<L_x$, $l_y=L_y$ and $l_z<L_z$. The typical dimensions of the graphene sheets studied in this work had lengths ($l_x$) 2.3-84.9 nm and widths ($l_y$) 6 nm, while the vacuum gap along the *x*- and *z-axis* was set to about 10 nm. The thickness of the SLGs was assumed to be equal to the interlayer distance of MLG, $l_z$=0.335 nm [49,50]. In order to avoid spurious interactions along the directions of *x*- and *z-axis* the differences $L_x-l_x$ and $L_z-l_z$ were larger than the cut off distance of the used potentials, hence our system was periodic only along the *y-axis* direction as it is illustrated in figure **1b**. In practice, this means that our results concern graphenes of relatively large widths. By replicating the graphene sheets along the *z-axis* direction, MLGs with up to 6 layers were constructed (as shown in figures **1c**, **1d** and **1e** for 2, 3 and 4 layers, respectively). It should be noted that the AB stacking was found to be stable in contrast to the case of AA stacking in accordance with Ref. [51].

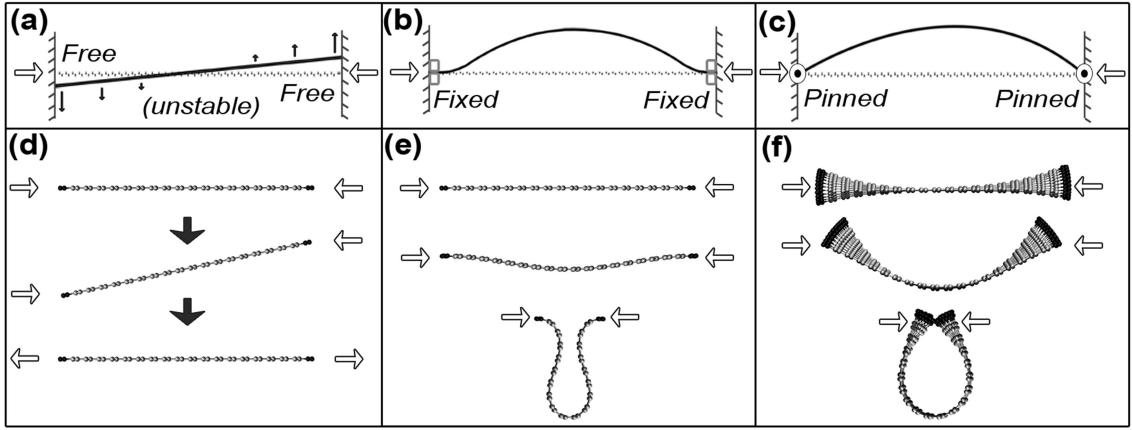

**Figure 2.** Schematic representations of loaded slabs with **(a)** free, **(b)** fixed (clamped) and **(c)** pinned (simply supported) edges. Configurations of **(d)** free, **(e)** fixed and **(f)** AMC-constrained graphenes under uniaxial compressive forces, as obtained from MD simulations. The white arrows display the direction of the constant applied forces. In (d) snapshots of the evolution (indicated by the black arrows) of the process that leads to the rotation of the SLG are shown. Equilibrium configurations for different loads are presented in (e) and (f): the applied forces are slightly smaller (top), slightly larger (middle) and much larger (bottom) than the critical buckling stress $\sigma_{crit}$.

The applied constrains at the loading edges of graphene play an important role to its mechanical response, as it has been demonstrated by continuum models [35,52]. In the case where no geometrical constraints are applied (free edges) the graphene sheets can simply rotate under compressive uniaxial loads, as it is illustrated in the schematic of the figure **2a** and the atomistic representation of a MD simulation in figure **2d**. One method to prevent rotations of the graphenes in MD is to constrain the edges of graphene in the *xy-plane* (fixed edges, see schematic of **2e**). The behavior of those systems is quite similar with the predictions of the continuum model for slabs with clamped (fixed) edges (see Fig. **2b**), since in both cases the displacement *w* of the edges along the *z-axis* and its derivative (*dw/dx*) equal to zero [35]. Another method implemented in lammps[38] to prevent rotations of a nanostructure is to maintain its angular momentum to zero, by subtracting at each time step of the simulation the angular velocity component of each atom with respect to the center of mass. In this case, the graphenes under the compressive forces were bended as shown in figure **2f**. This condition will be referred as "angular momentum conserving" (AMC) constrain. It should be noted that this approach cannot be applied effectively to ribbons where $l_y$ is much larger than $l_x$ since they are able to respond by twisting instead of bending (see for example the Fig. **S1** in the supplementary material information). AMC constrained graphenes feature a shape similar with that of pinned plates at the edges, in the sense that the displacement $w$ exhibits an almost linear behavior indicating that $d^2w/dx^2 \approx 0$. In plates with pinned edges the displacement $w$ and its second derivative ($d^2w/dx^2$) at their edges are zero (see Fig. **2c**). One strict difference is that in AMC the displacements $w$ of the edges are not zero. However, due to AMC and the symmetry of the structure, the relative displacement between the two edges of the graphene is zero, as happens with the case of pinned plates. Graphenes with fixed edges or under AMC constrains are considered here.

Initially, the graphenes free of any load were left to relax for 100 ps to achieve equilibrium at the desired temperature and pressure, and acquire the initial dimensions. Next, compressive forces of constant magnitude were applied to the atoms at the edges of graphene (black atoms in figure **1b**) for up to 10 ns. We have tested that this time is sufficient for achieving time independent results for the considered cases. The obtained strain was calculated by time averaging the value of the strain during the last 0.5 ns of the simulation. By performing MD simulations for a wide range of the applied forces, the stress-strain curves were extracted. We have checked that the depth of the region of the atoms that are constrained in-plane in the case of fixed edges, (see black atoms in **1b**, in this case the depth is one line of atoms) does not affect the response of the graphenes as long as the distance between those regions (the effective length of graphene) remains constant.

### III. Results and discussion

*a) Uniaxial compression of monolayer graphene*

Figure **3** displays the compressive strain-stress curves of SLGs with different lengths $l_x$ using the boundary condition of the fixed edges for the three different force fields. In the limit of small strains the Young's modulus (Poisson's ratio) was measured to ~0.93 TPa (0.2), 1.26 TPa (-0.1) and 0.86 TPa (0.3) for the LCBOP, Tersoff and REBO force fields, respectively.

The value of the obtained elastic modulus is more or less close to the experimental value of 1 TPa [6] for the LCBOP and REBO potentials, while the Tersoff overestimates it by 25%. The Poisson's ratio are also in agreement with other calculations [53–56]. The negative Poisson ratio obtained by the Tersoff potential is an artifact that has been reported in literature [57] and can be attributed to inaccuracies of the potential. As it can be seen from figure **3**, the behavior of the stress-strain curves is qualitatively similar for all the tested potentials. Above a critical value of the stress, the strain increases by more than one order of magnitude indicating the buckling of graphene. The critical buckling value depends on the length $l_x$ of graphene.

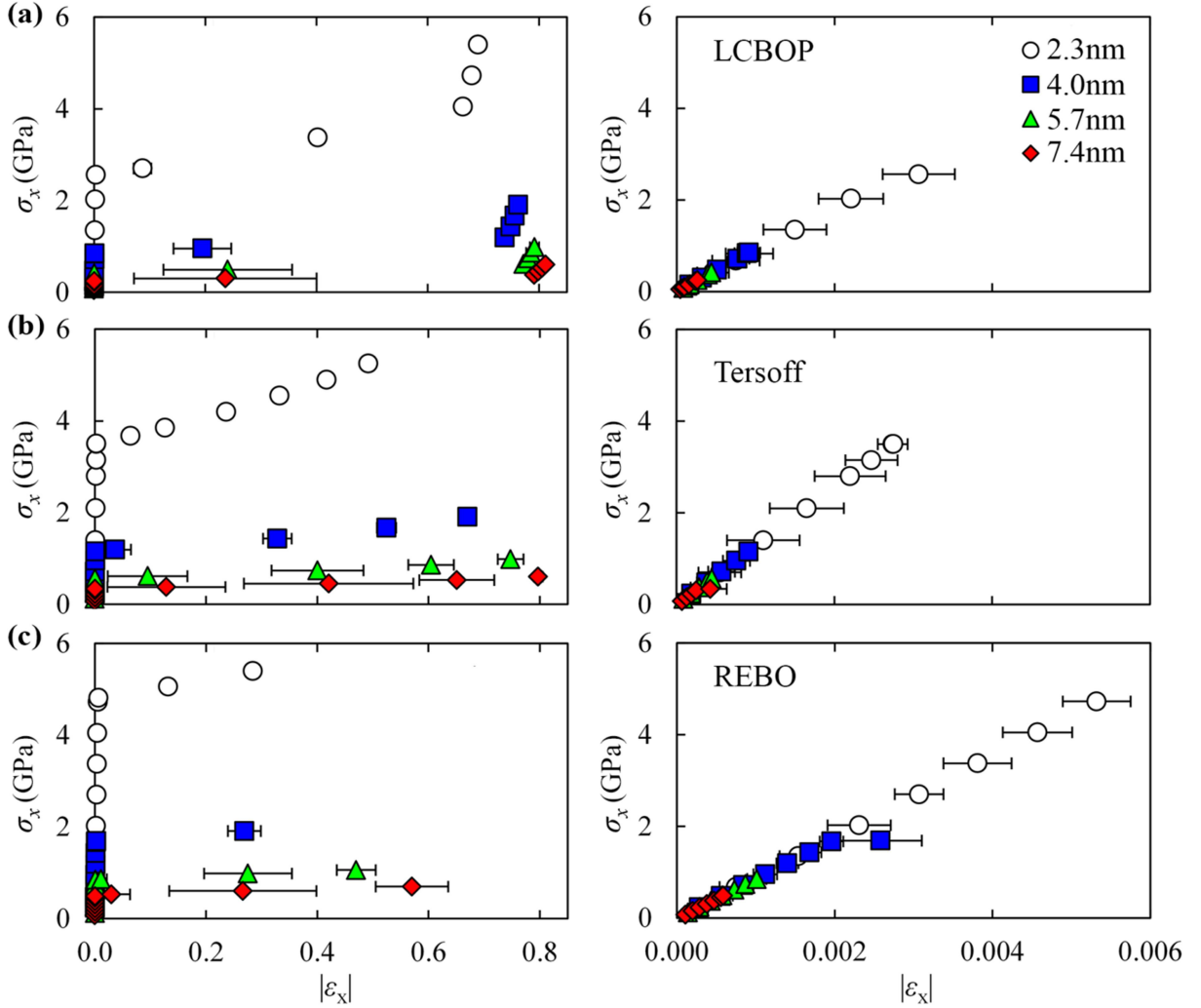

**Figure 3.** Compressive stress ($\sigma_x$) - strain ($|\varepsilon_x|$) diagrams for monolayer graphenes with fixed edges, using the **(a)** LCBOP, **(b)** Tersoff and **(c)** REBO force fields. Different lengths of graphene are shown, $l_x$=2.3 nm (circles), 4.0 nm (squares), 5.7 nm (triangles) and 7.4 nm (diamonds). The right column depicts an enlargement of the corresponding plots of low applied strains, showing the linear response. The error bars represent standard deviations of the obtained time-averaged strains.

Figure **4a** (**4b**) presents the critical buckling stress (strain) with respect to the length $l_x$ of graphene for the three force fields used. The critical buckling stress was calculated with a resolution lower than 0.01 GPa (the corresponding error bars in Fig. **4a** are negligibly small). The critical buckling strain equals to the value of strain just before the buckling instability

(see the Fig. **S2(a)** in the supplementary material). For all the used force fields, the length dependence of the critical values $\sigma_{crit}$ and $\varepsilon_{crit}$ are accurately described by an inverse square law:

$$\sigma_{crit} = \frac{\alpha}{l_x^2} \quad , \quad \varepsilon_{crit} = \frac{\alpha'}{l_x^2} \tag{1}$$

where $\alpha$ and $\alpha'$ are constants. This is in agreement with Refs. [30,32,33] where the critical buckling values were found to be inversely proportional to the length square. As we examined periodic systems in the direction of *y*-axis, simulations at different values of $l_y$ show no dependence of the critical buckling values on $l_y$, at least for the low temperatures considered here.

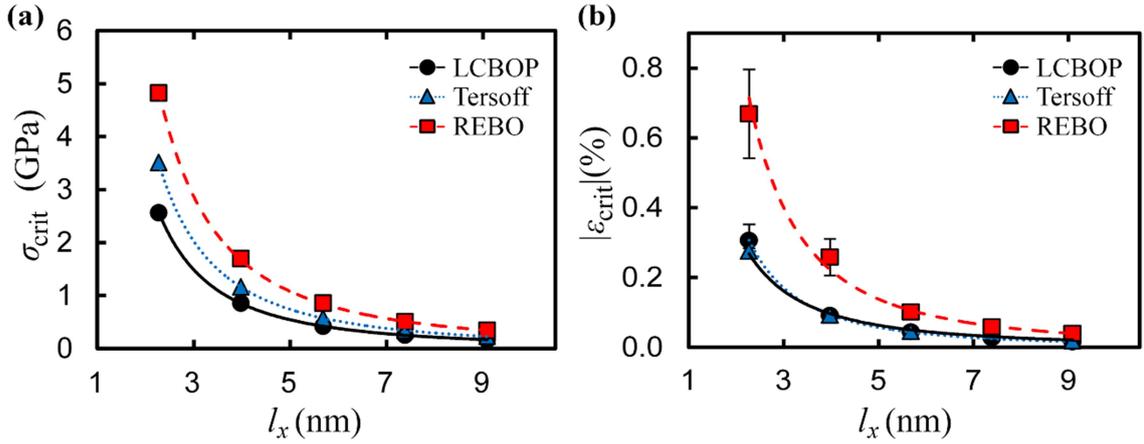

**Figure 4.** The critical buckling **(a)** stress and **(b)** strain as a function of the length $l_x$ of graphene using the LCBOP (circles), Tersoff (triangles) and REBO (squares) force fields. The error bars in (b) represent standard deviations of the time-averaged value. Lines are fittings of the numerical results with Eq.(1).

Qualitatively, the behavior of SLGs under the AMC constrain was quite similar with that of SLGs with fixed ends. This can be seen from Figs. **S3** and **S4** in the supplementary material information, which display the strain-stress diagram and the plots of $\sigma_{crit}$ and $\varepsilon_{crit}$ versus $l_x$, respectively, for this case using the LCBOP potential. However, the critical buckling stress was found to be about 3-4 times smaller than the corresponding value calculated for SLG with fixed ends, a result that is consistent with continuum models (see below).

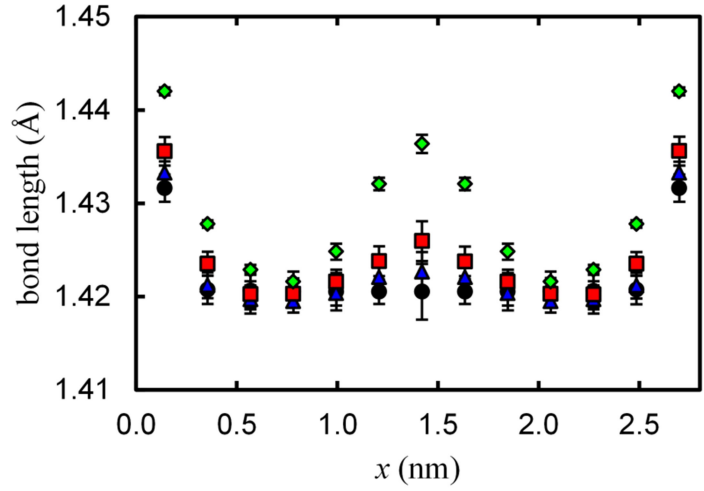

**Figure 5.** The distribution of the C-C bond lengths along the loading direction for graphene sheets with $l_x$=2.7 nm under compressive uniaxial loads equal to 0.0 (circles), 2.1 (triangles), 2.3 (squares) and 2.5 (diamonds) GPa. In this case $\sigma_{crit}$ = 1.8 GPa. The inset displays two graphene equilibrium configurations with $\sigma_x$= 2.1 GPa (top) and $\sigma_x$= 2.5 GPa (bottom). The error bars represent standard deviations of the time- and width-averaged values shown in the figure.

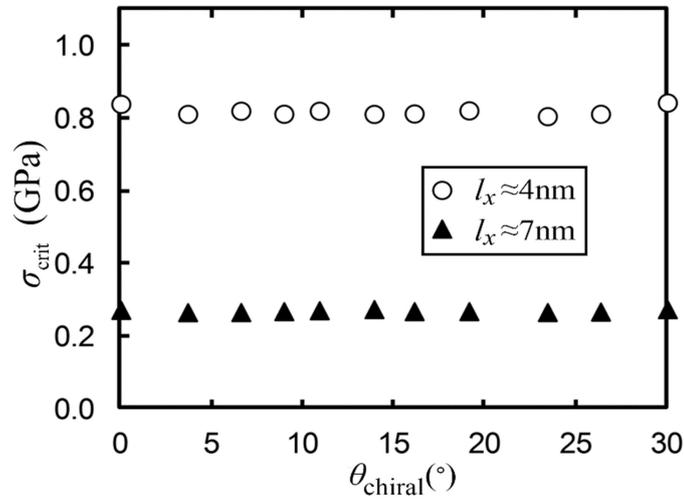

**Figure 6.** The critical buckling stress $\sigma_{crit}$ versus the chirality ($\theta_{chiral}$) of the loading direction for graphene sheets with $l_x\approx$4 nm (circles) and $l_x\approx$7 nm (triangles).

Figure **5** displays the distribution of the C-C bond lengths along the loading direction (*x-axis*) for a fixed-edged SLG with $l_x$=2.7 nm, for various values of the applied force above the critical buckling $\sigma_{crit}$, using the LCBOP force field. The C-C bonds in graphenes that are not subjected to compressive loads (Fig. **5**, black circles) are equal to 1.42Å in the central region of the sheet and about 1.6% larger at the boundary due to edge effects. However graphene sheets under loads that exceed the $\sigma_{crit}$ value (onset of buckling) show an increase of the bond length near the central region. This is in accordance with the observed relaxation of the shift of certain Raman bands after the onset of buckling of monolayer graphene [29,58].

In Refs. [30,33] the critical buckling strains ($\varepsilon_{crit}$) were found to be the same along the armchair and zigzag directions of the graphene. In the current study graphene sheets with loading direction at various angles were examined, in order to investigate the dependence of $\sigma_{crit}$ on the chiral angle ($\theta_{chiral}$) of the applied stress. It should be noted that the chiral angle can only take specific values in order to maintain proper periodicity of the lattice along the periodic boundaries of the simulation box. Figure **6** displays results for graphene sheets with $l_x \approx 4$ nm (circles) and $l_x \approx 7$ nm (triangles), uniaxially compressed at different chiral angles $\theta_{chiral}$ using the LCBOP potential. In both cases $l_y = 6$ nm. These results are in agreement with Ref. [30] where the REBO potential was used, and the $\sigma_{crit}$ was found independent from the chirality of the loading direction.

### b) Uniaxial compression of multilayer graphene

Here, the compressive response of multilayer graphenes is discussed, through MD simulations using the LCBOP force field. Figure **7** shows that the stress-strain curves of MLGs with fixed edges exhibit a similar behaviour with that of SLGs (see Fig. **3**). The buckling instability in this case is similar to that occurring in SLGs, as can be seen from Fig. **S2(b)** in the supplementary material section. In all presented cases, the Young's modulus was calculated in the regime 0.90-0.95 TPa. As excpected, for $\sigma > \sigma_{crit}$, there seems to be an increased resistance of the sheets for further bending as the number $N$ of layers increases. For example, for sheets with length $l_x = 2.3$ nm, when $\sigma_x = 4$ GPa the $\varepsilon_x$ of SLG is about 65% (derived from the data shown in Fig. **3a** with circles), while the corresponding one of MLGs with $N=2$, 3 and 4 layers is about $\varepsilon_x = 25\%$, 17% and 10% (derived from the data shown with circles in figures **7a**, **7b** and **7c**, respectively). Similar behaviour occurs for other values of $l_x$.

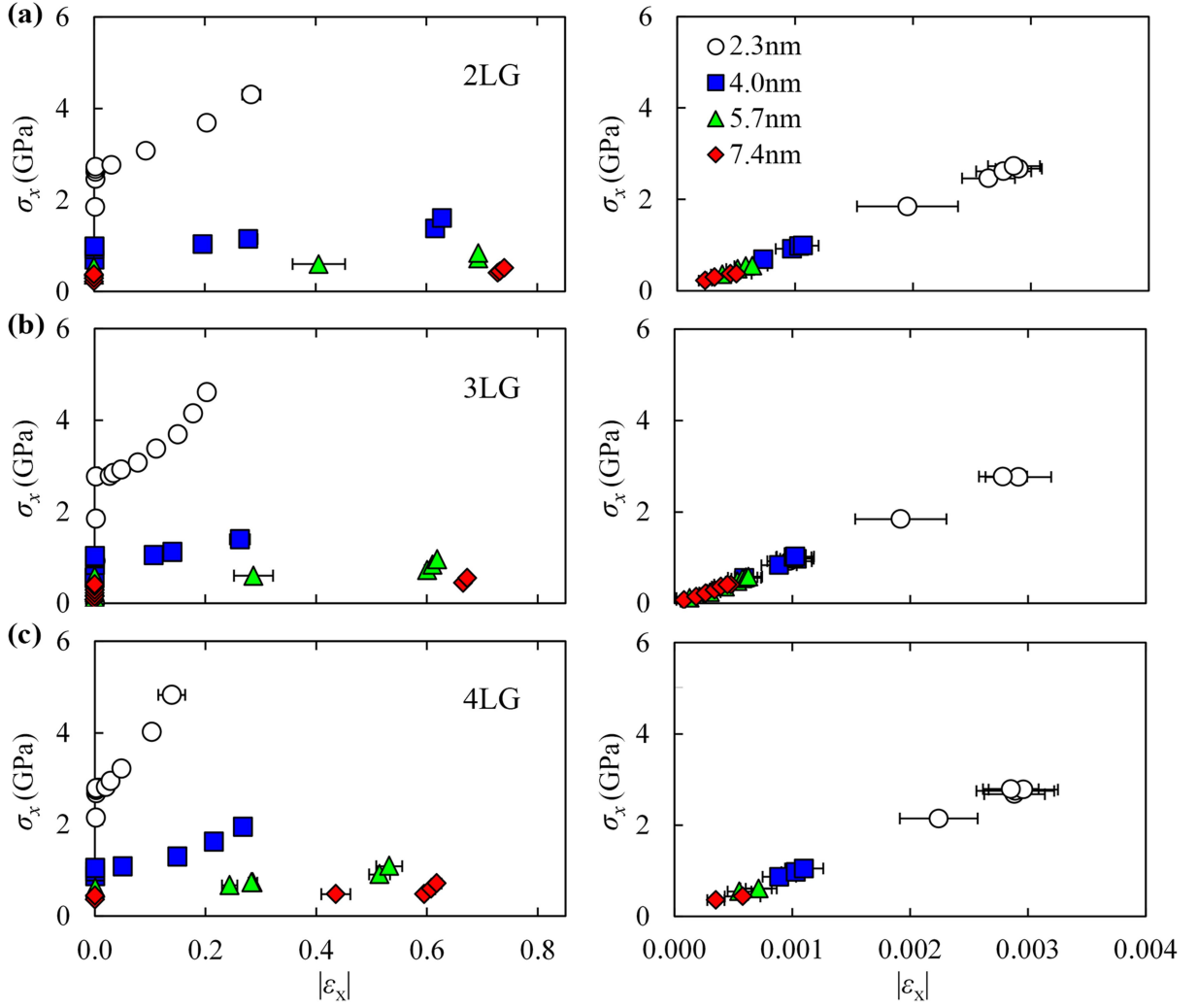

**Figure 7.** Compressive stress ($\sigma_x$) - strain ($|\varepsilon_x|$) diagrams for MLGs with fixed edges consisting of **(a)** two, **(b)** three and **(c)** four layers, using the LCBOP force field. Results for different lengths of MLGs are shown: $l_x$=2.3 nm (circles), 4.0 nm (squares), 5.7 nm (triangles) and 7.4 nm (diamonds). The right column depicts an enlargement of the corresponding plots of low applied strains, showing the linear response. The error bars represent standard deviations of the obtained time-averaged strains.

The critical buckling stress $\sigma_{\text{crit}}$ as a function of the length $l_x$ of MLGs with fixed edges and unsupported sides is presented in figure **8**, for $N$=2 up to $N$=6 layers. The SLG case with $N$=1 (circles in Fig. **4a**) is also shown for comparison. The solid line shows the fitting of SLG results with the inverse square law, Eq. (1). Figure **8a** clearly demonstrates that the critical buckling stresses of MLGs do not follow an inverse square law, but they exhibit a slower decrease that can not be described through a simple power law. Similar results have been obtained for bilayer graphenes ($N$=2) using the AIREBO potential as well (see figure **S5** in the supplementary material).

Multilayers under the AMC constrain displayed a similar behavior, as illustrated in Fig. **S6** of the supplementary information, regarding the $\sigma_{crit}$ dependence on $l_x$. Additionally, for AMC-constrained MLGs, the $\sigma_{crit}$ is about 2-4 times smaller than the corresponding $\sigma_{crit}$ value of MLGs with fixed ends.

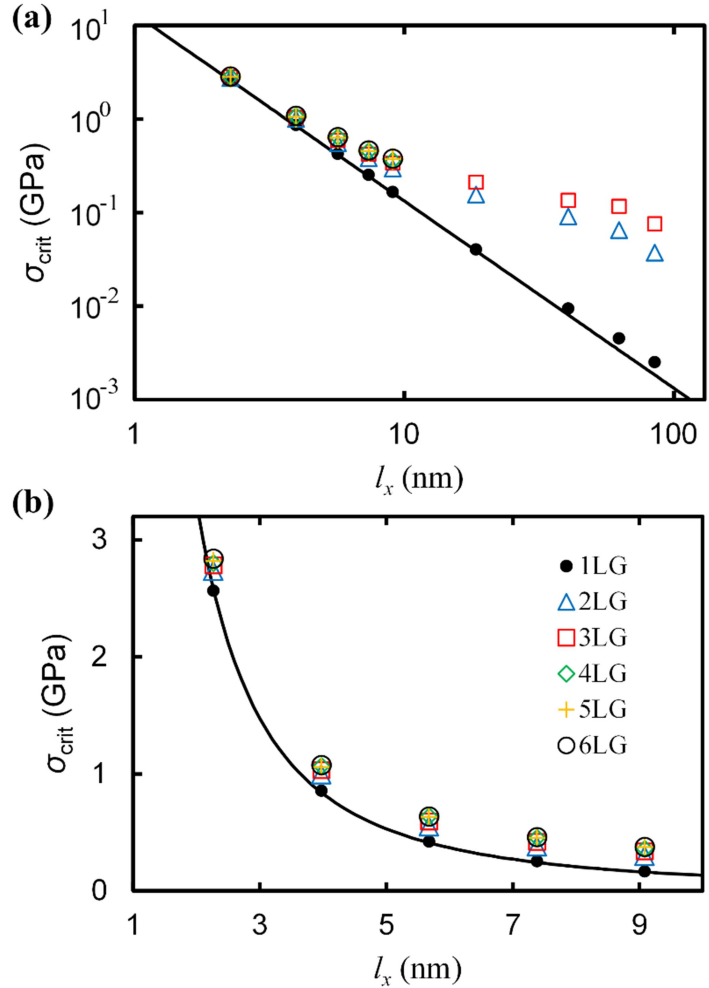

**Figure 8. (a)** Logarithmic plot of the critical buckling stress with respect to the length $l_x$ of *N*-layered graphenes with fixed edges, for *N*=1 (filled circles), *N*=2 (triangles), *N*=3 (squares), *N*=4 (diamonds), *N*=5 (crosses) and *N*=6 (open circles) layers, using the LCBOP force field. **(b)** The corresponding linear plot for graphene lengths up to 10 nm. Lines are fittings of the numerical results for the case of SLGs with an inverse square law, Eq (1).

c)  *Comparison with the linear elasticity theory for loaded slabs*

In the linear elasticity theory of continuum mechanics, the critical buckling stress of a wide plate with unsupported sides, length $l_x$, thickness $l_z$, Poisson's ratio $v$ and Young's modulus *E* is provided from the following relation [35]:

$$\sigma_{\text{crit}} = \frac{\pi^2}{12} \frac{E}{1-v^2} \frac{l_z^2}{(l_x/p)^2} = \frac{d}{(l_x/p)^2} \qquad (2)$$

where $p = 2$ for slabs with fixed edges, while $p = 1$ for slabs with pinned edges. This difference in the value of $p$ results in $\sigma_{\text{crit}}^{\text{fix}} = 4 \cdot \sigma_{\text{crit}}^{\text{pin}}$, were $\sigma_{\text{crit}}^{\text{fix}}$ ($\sigma_{\text{crit}}^{\text{pin}}$) is the critical buckling stress for plates with fixed (pinned) edges. Note that the Eq.(1) of Ref. [11] giving the critical buckling value for plates with simply supported sides, is reduced for wide plates to $1/l_x^2$ dependence of Eq. (2), taking into account that when the width $w$ is much larger than the length $l$, then the geometric term $K$ tends to $w^2/l^2$. Moreover, the same equation contains the thickness dependence ($\sim l_z^2$) appeared in Eq. (2), considering that the ratio $D/C$ of the flexural and tension rigidities equals to $h^2/12$ (see Eq. (2) of Ref. [28]), where $h$ is the thickness of the plate. For wide plates such a coincidence of formulae corresponding to different boundary conditions at the sides of the plate is expected, since in this case (where the width is very large) the boundaries at the sides play a negligible role.

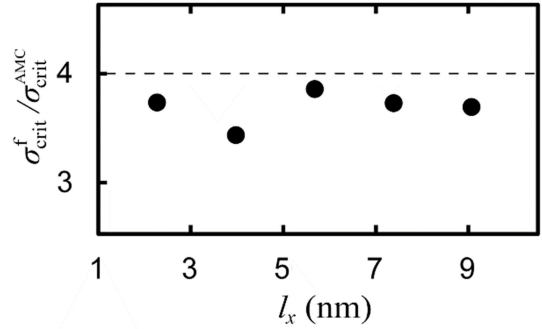

**Figure 9.** The ratio $\sigma_{\text{crit}}^{\text{f}}$ / $\sigma_{\text{crit}}^{\text{AMC}}$ (see text) for SLGs of various lengths ($l_x$). The dashed line displays the ratio of the critical buckling for fixed- and pinned- boundary conditions in the continuum model.

The results from the MD simulations in SLGs are in qualitative agreement with the continuum model since they give $\sigma_{\text{crit}} \propto l_x^{-2}$ [see Fig. **4a** and Eq. (1)]. However there is a discrepancy at a quantitative level: for a SLG with $E=1$TPa [6], $l_z=0.335$ nm [49] and $v=0.22$ [53], the constant $d$ in Eq.(2) equals to 97 GPa·nm$^2$, which is much larger than the value extracted from the MD simulations. The continuum model could be fitted to the results derived from the MD simulations by adjusting the conventional thickness of the SLG ($l_z=0.335$ nm) to an effective thickness that equals to about 0.04 nm. Such an unrealistic requirement for the graphene thickness, in order to fit atomistic results with the predictions of the continuum theory for plates, also appears in the case of graphene bending [59]. We also note that the ratio $\sigma_{\text{crit}}^{\text{f}} / \sigma_{\text{crit}}^{\text{AMC}}$ of the critical values for fixed-edged and AMC-constrained (showing some analogy with the behavior of pinned edges as discussed in section II) graphenes is close to 4 as it is displayed in figure **9**.

However, in the case of MLGs qualitatively different behaviors are obtained between the results of MD simulations and the continuum theory for loaded slabs with unsupported sides. In particular, as Fig. **8a** clearly shows, the MD calculations reveal significant deviations from the $1/l_x^2$ law of Eq. (2), concerning the length dependence of critical buckling values. Moreover, in the linear elasticity continuum theory the $\sigma_{\text{crit}}$ is proportional to the square of the thickness $l_z$ of the plate [see Eq. (2)], while the MD simulations show that $\sigma_{\text{crit}}$ only slightly increases with $l_z$, as can be seen from figure **8b**. Both fixed edged and AMC constrained graphenes reveal these features in the numerical simulations. These discrepancies between the MD results and the continuum theory may be due to the discrete nature and the interlayer interactions of MLGs.

## 4. Conclusions

The behavior of single layer and multilayer graphenes of large width under uniaxial compressive loads was investigated, through molecular dynamics simulations. The compressive tests were stress-controlled and the simulations took place in the isothermal-isobaric ensemble. The studied graphene sheets in this work had typical dimensions of length 2 nm < $l_x$ < 100 nm and thickness $l_z$≈0.3 - 2 nm (from one up to six graphene layers), while periodic boundary conditions were used along the $y$ direction. Different force fields and boundary conditions/constrains were used in our MD simulations. For SLGs the calculated elastic moduli were relatively close to the experimental value $E$ = 1 TPa, while the Poisson ratios of the LCBOP and REBO potentials were in the range 0.2-0.3, in agreement with values reported in the literature.

For single layer graphenes, the critical buckling stress/strain is inversely proportional to the square of the length of the graphene along the loading direction. Additionally, the $\sigma_{\text{crit}}$ was found to be independent on the loading direction (armchair, zigzag or any other chiral direction). The MD results for SLG show qualitative agreement with the continuum theory for uniaxial compressive deformation of wide plates of unsupported sides.

However, in the case of multilayer graphenes discrepancies were found between the MD calculations and the continuum linear elasticity theory. The latter one predicts that $\sigma_{crit}$ increases with the square of the thickness, while the MD simulations show a significantly smaller increase. Furthermore, the inverse squared length dependence of the critical buckling values, as anticipated in the continuum theory for plates, is not verified by the numerical simulations, where deviations from this behavior are obtained for MLGs with $N=2$ up to $N=6$ layers.

**Acknowledgments**

We would like to thank D. N. Theodorou and C. Tzoumanekas for valuable discussions. This work has been partially supported from the Thales project GRAPHENECOMP co-financed by the European Union (ESF) and the Greek Ministry of Education (through ΕΣΠΑ program), the ERC Advanced Grant 'Tailoring graphene to withstand large deformations'(Grant agreement:321124) financed by the European Research Council, and by European Union's Seventh Framework Programme (FP7-REGPOT-2012-2013-1) under grant agreement n° 316165. Financial support of Graphene FET Flagship (Graphene-Based Revolutions in ICT and Beyond, Grant No. 604391) is also acknowledged.